\documentclass[twocolumn,runningheads,natbib]{svjour2}
\smartqed  
\usepackage{graphicx}
\newcommand{\rxte}{RXTE}
\journalname{Astrophysics and Space Science}
\begin{document}

\title{New Phase-coherent Measurements of Pulsar Braking Indices}

\author{Margaret A. Livingstone         \and
        Victoria M. Kaspi \and
	Fotis P. Gavriil \and
	Richard N. Manchester \and
	E.V. Gotthelf \and
	Lucien Kuiper
}

\institute{M. Livingstone \at
              Physics Department, McGill University \\
	      Rutherford Physics Building\\
	      3600 University Street \\
	      Montreal, Qc H3A 2T8 \\
	      Canada \\
              \email{maggie@physics.mcgill.ca}           
           \and
           V.M. Kaspi \at
	   Physics Department, McGill University \\
	   Rutherford Physics Building\\
	   3600 University Street \\
	   Montreal, Qc H3A 2T8 \\
	   Canada 
	   \and
	   F.P. Gavriil \at
	   NASA Goddard Space Flight Center \\
	   Code 662, X-ray Astrophysics Laboratory \\
	   Greenbelt, MD 20771
	   \and
	   R.N. Manchester \at
	   Australia National Telescope Facility \\
	   CSIRO \\
	   P.O. Box 76 \\
	   Epping NSW 1710 \\
	   Australia 
	   \and
	   E.V. Gotthelf \at
	   Columbia Astrophysics Laboratory\\
	   Columbia University\\
	   550 West 120th Street \\
	   New York, NY \\
	   10027-6601 
	   \and
	   L. Kuiper \at
	   Netherlands Institute for Space Research\\
	   Sorbonnelaan 2 \\
	   3584 CA \\
	   Utrecht, Netherlands \\
}

\date{Received: date / Accepted: date}

\maketitle

\begin{abstract}
Pulsar braking indices offer insight into the physics that underlies
pulsar spin-down. Only five braking indices have been measured via 
phase-coherent timing; all measured values are less than 3, the value expected
from magnetic dipole radiation. Here we present new measurements for
three of the five
pulsar braking indices, obtained with phase-coherent timing for \\
PSRs~J1846$-$0258 ($n=2.65\pm0.01$), 
B1509$-$58 ($n=2.839\pm0.001$) and B0540$-$69 ($n=2.140\pm0.009$). 
We discuss the implications of these results
and possible physical explanations for them.
\end{abstract}

\section{Introduction}
\label{intro}
A very commonly assumed model for pulsar spin-down posits that 
\begin{equation}
\dot{\nu} = -K\nu^n,
\label{eqn:spindown}
\end{equation}
where $\nu$ is the pulse frequency, $\dot{\nu}$ is the frequency
derivative, $K$ is a constant and $n$ is the braking index. The braking
index is then given by
\begin{equation}
n=\frac{\nu \ddot{\nu}}{{\dot{\nu}}^2} .
\end{equation}
The braking index provides insight into the physics that
drives pulsar spin-down. Typically, it is assumed that magnetic dipole
radiation underlies pulsar evolution, resulting in $n=3$ \citep[e.g.][]{mt77}.
However, other processes could, in 
principle, cause the pulsar to radiate and would result in different values
for $n$ and $K$. For example, a pulsar spun down entirely by the loss of
relativistic particles would have $n=1$ \citep{mt69}. A pulsar losing energy via
gravitational radiation or quadrupole magnetic radiation would spin down
with $n=5$ \citep{br88}. 

Braking indices have proven difficult to measure. To date, only six have been
reported even though more than 1600 pulsars are known.  Evidently, the pulsar properties
necessary for a measurement of $n$ are rare; the pulsar must:
spin down quickly; experience few, small, and relatively infrequent
glitches; and be relatively uncontaminated by timing noise, a low-frequency stochastic process
superposed on the deterministic spin-down of the pulsar. The
youngest pulsars, of which the Crab pulsar is the most famous example,
uniquely possess these three qualities. The reason then for the paucity of
measured values of $n$ is a direct consequence of the relative rarity of
very young pulsars, i.e. those with characteristic ages on the order of
1\,kyr, where characteristic age is defined as
\begin{equation}
\tau_c \equiv \frac{P}{2\dot{P}} = \frac{\nu}{2\dot{\nu}}.
\label{eqn:age}
\end{equation}

Of the six pulsars with measured $n$, five were obtained via phase-coherent
timing. All five of these pulsars: PSRs~J1846$-$0258, B0531+21 (the Crab
pulsar, $n=2.51\pm0.01$), 
B1509$-$58, J1119$-$6127 ($n=2.91\pm0.05$), and B0540$-$69, have 
characteristic ages less than 2\,kyr \citep{lkgk06,lps93,lkgm05,ckl+00,lkg05}.
The sixth measurement, that of the Vela pulsar ($n=1.4\pm0.2$),
could not be obtained
with phase-coherent timing due to large glitches \citep{lpgc96}. Timing
noise and large glitches begin to seriously contaminate measurements
of $n$ when pulsars have characteristic ages $\sim5$\,kyr \citep{ml90,
mgm+04}. 

In this paper we report on long-term {\textit{Rossi X-ray Timing Explorer}} 
observations of three young pulsars, 
PSRs~J1846$-$0258, B1509$-$58 and B0540$-$69. We present braking index
measurements for each of these pulsars obtained via phase-coherent timing.

\section{Phase-coherent pulsar timing}
The most accurate method of extracting pulsar timing parameters is 
phase-coherent timing, that is, accounting for every turn of the pulsar. 
Pulse times of arrival (TOAs) are measured
and fitted to a Taylor expansion of pulse phase, $\phi$
at time $t$ given by
\begin{equation}
\phi(t)=\phi(t_0) + \nu_0(t-t_0) + 
\frac{1}{2}{\dot{\nu}_0}(t-t_0)^2+\frac{1}{6}{\ddot{\nu}}_0(t-t_0)^3 + ...,
\label{eqn:taylor}
\end{equation}
where subscript 0 denotes a parameter at the reference epoch, $t_0$. TOAs
and initial spin parameters are input to pulse timing software (e.g.
TEMPO\footnote{http://www.atnf.csiro.au/research/pulsar/tempo}) 
and refined spin parameters and timing residuals are output.

The existence of timing noise and glitches in young pulsars is well known
to contaminate the measurement of deterministic spin parameters. Though powerful, 
a fully phase-coherent timing solution can be sensitive to 
these contaminants. In such cases, a partially coherent method may be
employed. In this case, local phase-coherent measurements of $\nu$, $\dot{\nu}$ and
possibly $\ddot{\nu}$ are made. Though the effects of timing
noise cannot be eliminated, the noise component is more readily identified
and separated from the deterministic component of the spin-down, 
as shown in Section~\ref{sec:1509}. In addition,
glitches can be easier to identify with this method, 
as will be shown in Section~\ref{sec:0540} of this paper. 

\section{Observations}
\label{sec:obs}
In this paper we describe observations of three young pulsars
PSRs~J1846$-$0258, B1509$-$58, and B0540$-$69 taken with the
Proportional Counter Array (PCA) on board the \textit{Rossi X-ray Timing
Explorer} (\rxte). The PCA consists of five collimated
xenon/methane multianode proportional counter units (PCUs). The PCA operates 
in the 2-60\,keV energy range, has an effective area of 
$\sim$6500\,cm$^2$ and has a 1 degree field of view. 
While \rxte\ has no imaging capability, it has excellent time resolution
of $\sim 1$\,$\mu$s \citep{jsg+96}. This makes \rxte\ ideal for observing
young, rapidly rotating pulsars. 

Observations of PSR~B1509$-$58 were taken in \\
``GoodXenonWithPropane'' mode,
while observations of the other two sources were taken in ``GoodXenon'' mode.
Both modes record the photon arrival time with $1\mu$s-resolution and photon
energy with 256-channel resolution. The number of PCUs active during an
observation
varies, but is typically three. For PSRs~B1509$-$58 and J1846$-$0258, 
which have relatively hard spectra, all three Xenon layers and photons
with energies ranging from 2-60\,keV were used, while for
the softer spectrum source, PSR~B0540$-$69, only the top Xenon layer and
photons
with energies ranging from 2-18\,keV were used. Further details of X-ray and radio
observations of 
PSR~B1509$-$58 are given in \cite{lkgm05} and references therein.
Details of \rxte\ observations of PSR~B0540$-$69 can be found in \cite{lkg05}
while details of PSR~J1846$-$0258 observations can be found in \cite{lkgk06}
and references therein.

Data were reduced using standard FITS tools as well as in-house software
developed for analyzing \rxte\ data for pulsar timing. Data from different
PCUs were merged and binned at (1/1024)\,s resolution. Photon arrival times
were corrected to barycentric dynamical time (TDB) at the solar system 
barycenter using the J2000 source positions and the 
JPL DE200 solar system ephemeris. 

For PSRs~B0540$-$69 and J1846$-$0258, initial ephemerides were found by
performing
periodograms on observations to determine values of $\nu$. Several values of $\nu$
were fitted with a linear least squares fit to determine an initial value of 
$\dot{\nu}$. These initial values were then used as input to a Taylor 
expansion of TOAs to determine more accurate parameters 
(Eq.~\ref{eqn:taylor}). PSR~B1509$-$58 has a previously determined ephemeris from radio timing data
obtained with the Molonglo Observatory Synthesis Telescope and 
the Parkes Radio Telescope \citep{kms+94}. We were able to extend that
fit with 7.6\,yr of \rxte\ data, removing a constant, but not well
determined, offset between radio and X-ray TOAs.

\section{PSR J1846$-$0258}
\label{sec:1846}
PSR J1846$-$0258 is a very young pulsar ($\tau_c=723$\,yr) located at the
center of the supernova remnant Kesteven 75. It has a relatively long spin
period of 324\,ms and a large magnetic field\footnote{$B \equiv 3.2 \times
10^{19} (P\dot{P})^{1/2}$G} of $B \sim 5 \times
10^{13}$\,G. PSR~J1846$-$0258 has been observed with \rxte\ for 6.3\,yr since 
its discovery in 1999 \citep{gvb+00}.

\begin{figure}
\includegraphics[width=0.5\textwidth]{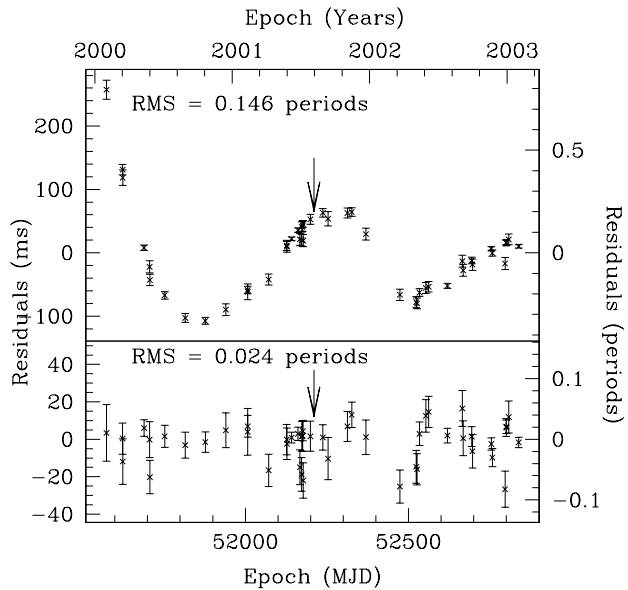}
\caption{Phase-coherent X-ray timing analysis of the young pulsar 
PSR~J1846$-$0258 spanning a 3.5-yr interval in the range MJD~51574-52837 (after
\cite{lkgk06}). Top panel: Residuals with $\nu$, $\dot{\nu}$, $\ddot{\nu}$ 
as well as glitch
parameters $\Delta{\nu}$ and $\Delta{\dot{\nu}}$ fitted. The glitch epoch,
MJD~52210 is indicated by the arrow. Bottom panel: Residuals with 
glitch parameters and eight frequency derivatives in total fitted
to render the residuals consistent with Gaussian noise. }
\label{fig:kes1}
\end{figure}

Using our initial ephemeris we obtained a \\
phase-coherent timing solution valid over a 3.5\,yr interval in the range 
MJD 51574-52837. Three spin parameters ($\nu$, $\dot{\nu}$ and $\ddot{\nu}$)
were required by the fit. In addition, we discovered
a small glitch near MJD 52210$\pm$10. The fitted glitch parameters are 
$\Delta{\nu}/{\nu}=2.5(2) \times 10^{-9}$ and
$\Delta{\dot{\nu}}/{\dot{\nu}}=9.3(1)\times10^{-4}$. 
Note that these and all other quoted uncertainties are 68\% confidence
intervals, unless otherwise indicated. The wide 
spacing of data near the glitch prevent the detection
of any short-timescale glitch recovery. Timing residuals are shown
in Figure~\ref{fig:kes1}. The top panel of Figure~\ref{fig:kes1} shows 
residuals 
with $\nu$, $\dot{\nu}$, $\ddot{\nu}$ and glitch parameters fitted. The residuals
clearly show systematics due to timing noise and possibly unmodeled glitch 
recovery. In order to minimize contamination of long-term timing parameters, 
we fitted additional frequency derivatives to render the residuals consistent
with Gaussian distributed residuals 
\citep[a process known as `whitening' residuals. See, for example, ][]{kms+94}. For this timing
solution, a total of eight frequency
derivatives were fitted, shown in the bottom panel of Figure~\ref{fig:kes1}.
The braking index resulting from this `whitened' timing solution is 
$n=2.64\pm0.01$. Complete spin-down parameters are given in Table~\ref{table:kes}.

\begin{table}[t]
\caption{Spin parameters for PSR~J1846$-$0258}
\centering
\label{table:kes}
\begin{tabular}{lll}
\hline\noalign{\smallskip}
Parameter & First solution & Second solution \\[3pt]
\tableheadseprule\noalign{\smallskip}
Dates (MJD)            & 51574.2 - 52837.4 & 52915.8 - 53578.6 \\
Epoch (MJD)            & 52064.0 & 53404.0 \\
$\nu$ (s$^{-1}$)      & 3.0782148166(9)&3.070458592(1) \\
$\dot{\nu}$ ($10^{-11}$s$^{-2}$) & $-$6.71563(1)&$-$6.67793(5) \\
$\ddot{\nu}$ ($10^{-21}$s$^{-3}$) &3.87(2) & 3.89(4)\\
Braking Index ($n$)    & 2.64(1)& 2.68(3)\\
Glitch epoch (MJD)     & 52210(10)& \\
$\Delta{\nu}/{\nu}$     & $2.5(2)\times10^{-9}$ & \\
$\Delta{\dot{\nu}}/{\dot{\nu}}$ & $9.3(1)\times10^{-4}$ & \\
\noalign{\smallskip}\hline
\tableheadseprule\noalign{\smallskip}
\noalign{\smallskip}\hline
\end{tabular}
\end{table}

Phase was lost over a 78-day gap in the data near MJD~52837, 
indicated by the fact
that a timing solution attempting to connect over this gap fails to predict the pulse 
frequency at previous epochs. The loss of phase is likely due to timing noise
or a second glitch. However, these two possibilities cannot be distinguished
due to the relatively long gap in the data set. 

\begin{figure}
\includegraphics[width=0.5\textwidth]{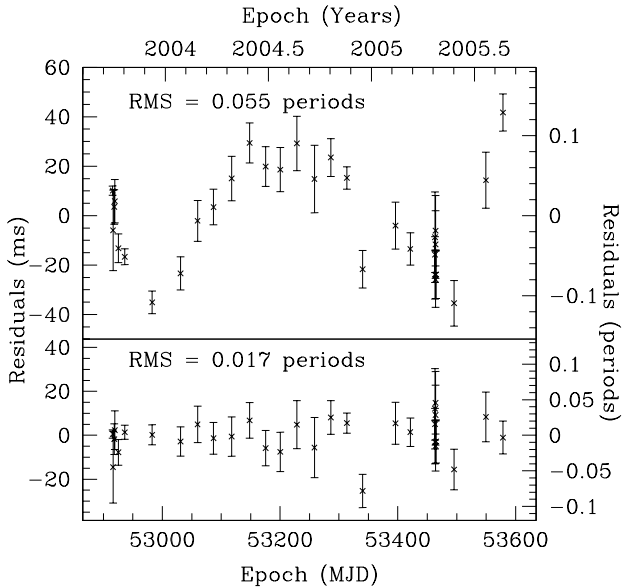}
\caption{Phase-coherent X-ray timing analysis of PSR~J1846$-$0258 spanning
an 1.8-yr interval from MJD~52915-53579 (after
\cite{lkgk06}). Top panel: Residuals with
$\nu$, $\dot{\nu}$, $\ddot{\nu}$ fitted. Bottom panel: Residuals with five 
frequency derivatives total fitted to render the residuals consistent with
Gaussian noise.}
\label{fig:kes2}
\end{figure}

A second phase-coherent solution was obtained for 1.8\,yr from MJD~52915-53579
with $\nu$, $\dot{\nu}$ and $\ddot{\nu}$. Timing residuals with these 
three parameters fit are shown in the top panel of Figure~\ref{fig:kes2}.
Again, systematics due to timing noise and/or glitch recovery remain in these
residuals. To `whiten' timing residuals, five total frequency derivatives
were fitted from the data, shown in the bottom panel of Figure~\ref{fig:kes2}.
Complete spin parameters for this timing solution are given in Table~\ref{table:kes}. 

Quoting the average value of $n$ from the two independent timing solutions,
which are in agreement, 
gives $n=2.65\pm0.01$. As is the case for all measured values of $n$, this
value is significantly less than 3, the value consistent with spin-down via
magnetic dipole radiation. This implies that some other physical process
must contribute to the spin-down of all of these pulsars. 

This measurement of $n$ for PSR~J1846$-$0258 
increases its age estimate \citep{lkgk06}. The commonly known characteristic age
(Eq.~\ref{eqn:age}) implicitly assumes that $n=3$.
A more physical estimate can be made once $n$ is known. The
age then can be estimated as 
\begin{equation}
\tau = \frac{1}{n-1} \tau_c \le 884\, {\rm yr}.
\label{eqn:nage}
\end{equation}
The estimate is an upper limit since the initial spin frequency of the
pulsar is not known. The upper limit approaches an equality when the pulsar
is born spinning much faster than its present spin frequency. 
Given the long period of the pulsar and the estimated initial spin period
distribution \citep[e.g. ][]{fk06}, the latter is likely to have occurred.
This age estimate for PSR~J1846$-$0258 is less than the known age of the 
Crab pulsar of 952\,yr.

\section{PSR~B1509$-$58}
\label{sec:1509}
\begin{figure}
\includegraphics[width=0.5\textwidth]{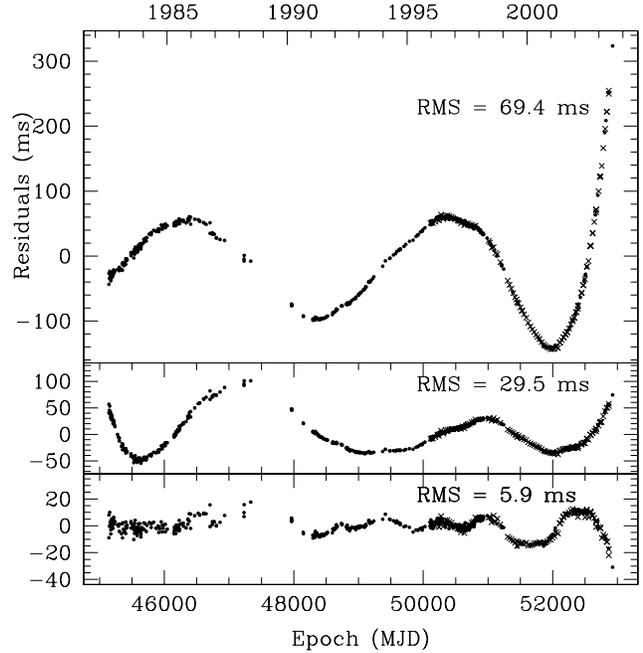}
\caption{Timing residuals for PSR~B1509$-$58. Radio TOAs are shown as dots
and the X-ray TOAs are shown as crosses (after \cite{lkgm05}). The top panel has pulse frequency
and three frequency derivatives removed, the middle panel also has the
fourth frequency derivative removed, and the bottom panel shows residuals
after the removal of five frequency derivatives.}
\label{fig:1509resids}
\end{figure}

The young pulsar PSR~B1509$-$58 was discovered in 1982 and has been observed
regularly ever since, first with radio telescopes such as the Molonglo
Observatory Synthesis Telescope \citep{mdn85} and the Parkes Radio Observatory
 \citep{kms+94}, and more recently with \rxte\ \citep{arn04,lkgm05}.
We phase-connected all 21.3\,yr of available radio and X-ray timing data to determine the
braking index. Timing residuals are shown in Figure~\ref{fig:1509resids}. The top panel 
shows the timing
residuals with $\nu$ and three frequency derivatives fitted; the middle
panel shows residuals with the fourth frequency derivative also fitted; the bottom
panel shows timing residuals with five frequency derivatives
fitted. Remarkably for such a young pulsar, \textit{no} glitches were
detected in this time period. Spin parameters from this phase-coherent
analysis are \\ 
$\nu = 6.633598804(3)$\,s$^{-1}$,
$\dot{\nu}=-6.75801754(4)\times 10^{-11}$\,s$^{-2}$, $\ddot{\nu}=1.95671(2)\times 10^{-21}$\,s$^{-1}$ at epoch MJD~49034.5.
These parameters imply a braking index of $n=2.84209(3)$, though
timing noise that could not be completely removed by fitting additional
frequency derivatives contributes to a systematic uncertainty
that is not included in the formal uncertainty quoted here. To solve this
problem, we performed a partially phase-coherent analysis by making 
independent measurements of $n$. 

\begin{figure}
\includegraphics[width=0.5\textwidth]{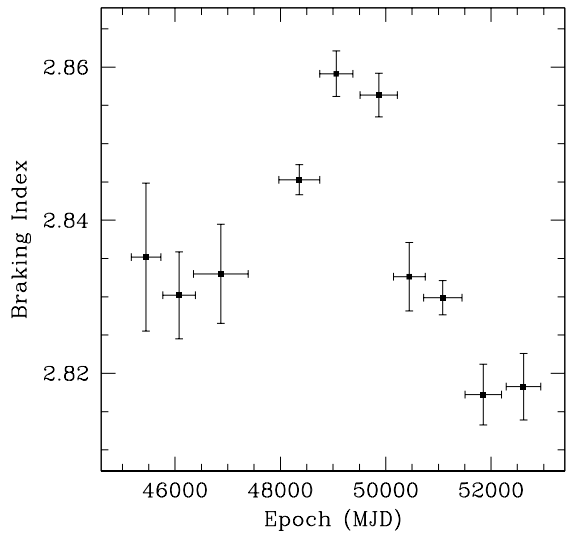}
\caption{Braking index calculated at 10 epochs of $\sim$2\,yr in length
(after \cite{lkgm05}). There is no statistically significant secular change 
of 21.3\,yr of data. The
average value is $2.839\pm0.003$, in agreement with the previously reported value
\citep{kms+94} and the value obtained from a phase-coherent analysis.
The reduced $\chi^2$ value is 15 for 9 degrees of freedom, suggesting
contamination by timing noise. }
\label{fig:1509n}
\end{figure}

Due to the large value of $\dot{\nu}$ for this pulsar, a significant
measurement of $n$ can be made in approximately 2\,yr, without noticeable
contamination of the measured spin parameters from timing noise. Thus,
having over 20 years of data allows 10 independent measurements of $n$,
which are shown in Figure~\ref{fig:1509n}. No secular variation of $n$ 
over 21.3\,yr is
seen, however, there is significant deviation from the average value of
$n=2.839\pm0.003$. This uncertainty was determined by a `bootstrap'
analysis which is a robust method of determining
uncertainties when the formal uncertainties are thought to underestimate
the true values, i.e. due to the presence of timing noise \citep{efr79}. 
Note that this value is in agreement with that obtained
with the fully phase-coherent timing solution, as well as
the previously reported value of $n=2.837 \pm 0.001$
\citep{kms+94}. The reduced $\chi^2$ is 15 for 9 degrees of freedom. 
This variation is likely due to the same timing noise
process that can be observed in timing residuals. Here, the variation is at
the $\sim1.5\%$ level. A similar analysis has been performed for 
PSR~J1846$-$0258
where variations are seen to be on the order of $\sim5\%$, though only at
the $2\sigma$ level \citep{lkgk06}
and for the Crab pulsar where variations are on the order of $0.5\%$ 
\citep{lps93}.

\section{PSR~B0540$-$69}
\label{sec:0540}
Located in the Large Magellenic Cloud, PSR~B0540$-$69 is commonly known as
the `Crab Twin', due to its similar spin and nebular properties. For
instance, its period of 50\,ms and magnetic field 
$B\sim5\times10^{12}$\,G are nearer to those of the Crab
pulsar ($P=33$\,ms, $B\sim4\times10^{12}$\,G) than for any other
pulsar. Due to its large distance, PSR~B0540$-$69 is very difficult
to detect in the radio waveband \citep{mml+93}, hence regular radio timing
of this source is not practical. 

The lack of regular, long-term timing observations for this pulsar has led to
conflicting values of $n$ in the literature; reported values range from 
$n=1.81\pm0.07$ to $n=2.74\pm0.01$ \citep{zmg+01,oh90}. 
Widely spaced timing observations greatly increases the risk of losing phase
if a phase-coherent solution is attempted. If instead of a phase-coherent
timing solution, measurements of frequency are obtained over widely spaced
intervals, small glitches can easily be missed and the effects of timing 
noise are difficult to discern. Two conflicting values of $n$ are of
particular interest since they are based on overlapping data from \rxte.

\cite{zmg+01} reported on 1.2 years of regular timing observations and found
a small magnitude glitch at MJD 51325$\pm$45
with parameters $\Delta{\nu}/{\nu} = (1.90 \pm 0.04) \times 10^{-9}$ and
$\Delta{\dot{\nu}}/{\dot{\nu}} = (8.5 \pm 0.5) \times 10^{-5}$. They used
the 300 days of data available after the glitch to measure a braking
index of $n=1.81\pm0.07$.  \cite{cmm03} extended the data set to 4.6\,yr and
reported that no glitch occurred. In contrast to 
the previous value, they measured $n=2.125\pm0.001$.

\begin{figure}
\includegraphics[width=0.5\textwidth]{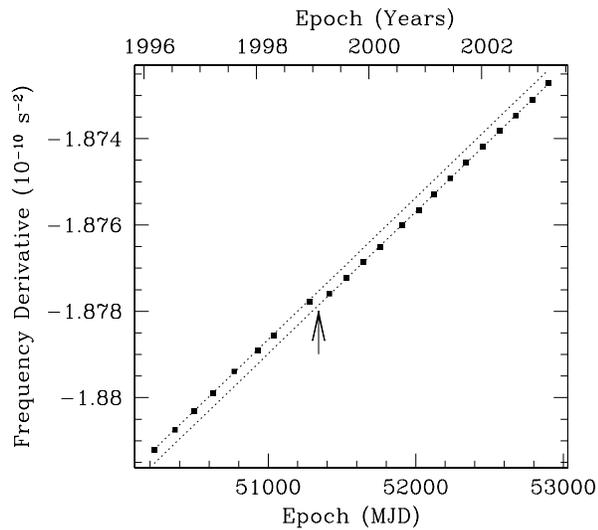}
\caption{Measurements of $\dot{\nu}$; the slope is $\ddot{\nu}$ (after
\cite{lkg05}). The
glitch occurring near MJD~51342 is shown with an arrow. The pre-glitch slope
is $\ddot{\nu} = 3.81(3) \times 10^{-21}$s$^{-3}$, while the post-glitch
slope is $\ddot{\nu}=3.81(1) \times 10^{-21}$s$^{-3}$. The average of pre- and
post-glitch $n$ is $2.140\pm0.009$. Measurement uncertainties are smaller 
than the points. }
\label{fig:0540}
\end{figure}

We re-examined all previously reported \rxte\ data and extended the data
set by 3\,yr in order to resolve the discrepant timing solutions and measure
the true braking index for this source. We phase connected a total of
7.6\,yr of data and found a small glitch near MJD~51335 with parameters 
$\Delta{\nu}/{\nu}
\sim 1.4 \times 10^{-9}$ and $\Delta{\dot{\nu}}/{\dot{\nu}} \sim 1.33 \times
10^{-4}$, in agreement with those reported by \citet{zmg+01}. 
This glitch is very small, and is most easily seen by the change in
$\dot{\nu}$ at the glitch epoch. Figure~\ref{fig:0540} shows
22 measurements of $\dot{\nu}$ obtained from individual phase-coherent
analyses, with the fitted glitch epoch indicated by an arrow. The
slope of the line, that is, the second frequency derivative $\ddot{\nu}$,
does not change significantly after the glitch (before  $\ddot{\nu} =
3.81(3)\times10^{-21}$s$^{-3}$,  after
$\ddot{\nu}=3.81(1)\times10^{-21}$s$^{-3}$). Uncertainties on $\ddot{\nu}$
were determined by a bootstrap analysis \citep{efr79}. 
We use the average to determine the braking index, found to 
be $n=2.140\pm0.009$.

In agreement with \cite{zmg+01}, we report a small glitch near MJD~51335,
though our value of $n$ is significantly larger. By phase-connecting only
the same 300\,day subset of data that they used to measure $n$, we find
$n=1.82\pm0.01$, in agreement with their result. The low value of $n$
in this case appears to be the result of timing noise and/or glitch recovery 
contaminating the relatively short time baseline used to measure $n$. 

Our measured value of $n$ is $1.7\sigma$ from that reported by
\citet[][$n=2.125\pm0.001$]{cmm03} though they do not report a glitch and
their uncertainty does not account for the effects of timing noise. The
reason for the agreements between our measured values is that their value
of $n$ was determined by two phase-coherent fits to the data, before
and after the glitch epoch reported by \cite{zmg+01}, despite the fact that
\cite{cmm03} report no glitch. 

\section{Implications and Physical Explanations for $n<3$}
All measured values of $n$ are less than 3, the value consistent with spin-down due
solely to magnetic dipole radiation. This implies that an additional torque is
contributing to the spin-down of young pulsars. Also intriguing is the
relatively wide range of measured values of $n$, shown in Table~\ref{tab:index}.
A measurement of $n$ immediately provides a correction to the age estimate
of the pulsar given by the characteristic age, as shown with
Equation~\ref{eqn:nage}. Comparisons of the characteristic age and age
estimated with $n$ are also given in Table~\ref{tab:index}. Although the age
estimate is always increased with a measurement of $n<3$, it should be noted that the age estimates
given are upper limits since the initial spin frequencies are not
known. The calculation of magnetic fields are also affected by a measurement
of $n<3$, since these are obtained assuming pure magnetic dipole radiation.
Unfortunately, there is no simple formula to estimate the correction to the
dipole magnetic field as there is for the age. Specific details of the
spin-down torque are required to uncover the true magnetic field of pulsars. 
\begin{table}[t]
\caption{Braking index measurements via phase-coherent timing. Also given are
the characteristic age, $\tau_c$ and age estimate using $n$, $\tau$.}
\centering
\label{tab:index}       
\begin{tabular}{lccc}
\hline\noalign{\smallskip}
Pulsar & $n$ & $\tau_c$ & $\tau$  \\[3pt]
\tableheadseprule\noalign{\smallskip}
J1846$-$0258 & 2.65(1)  &  723 & 884 \\
B0531+21     & 2.51(1)  & 1240 & 1640 \\
B1509$-$58   & 2.839(3) & 1550 & 1690 \\
J1119$-$6127   & 2.91(5) & 1610 & 1680 \\
B0540$-$69    & 2.140(9) & 1670& 2940 \\
\noalign{\smallskip}\hline
\end{tabular}
\end{table}

There are several theories that attempt to explain the measurements of $n<3$. One
explanation is that the pulsar's magnetic field grows or counter-aligns 
with the spin axis. 
This is equivalent to allowing the `constant', $K$, in the simple model of pulsar
spin down (Eq.~\ref{eqn:spindown}) to vary with time \citep{br88}. An 
advantage of this model is that it can be tested if precision measurements 
of the third frequency derivative can be made \citep{bla94}. To date, 
the third frequency
derivative has been measured only for the Crab pulsar \citep{lps93} and 
PSR~B1509$-$58 \citep{lkgm05}, though neither is known with sufficient
precision
to rule out the null hypothesis of a constant value of $K$. Timing noise and in the
case of the Crab pulsar, glitches, may prevent a sufficiently precise measurement
from ever being made. 

Another suggestion is that a fall-back disk formed from supernova material 
modulates the spin-down of young pulsars, providing a propeller torque in
addition to the torque from magnetic dipole radiation. This would cause the 
pulsar to 
lose energy more quickly leading to a measured value
$2<n<3$ \citep{aay01}. A difficulty in this model is that the 
disk must not suppress the pulsed radio emission during the propeller phase 
\citep{mph01}. 

In recent years, much work has been done on modelling the pulsar 
magnetosphere. Fully physical, three dimensional, time-dependent 
models of the pulsar magnetosphere
are still some time away, however, significant progress has been made and
there is some suggestion that $n<3$ may be a natural result of a plasma filled
magnetosphere \citep[see, for example][]{spit05,tim06, cs06}.

The idea that plasma in the magnetosphere affects the torque acting on a
pulsar is gaining acceptance with the first observational evidence for this
having recently
being presented. \cite{klo+06} show that \\
PSR~B1931+24, which
has curious quasi-periodic nulling behaviour, spins down at different rates
when it is `on' or `off'. Specifically, the pulsar has a faster rate of spin down when
it is observed in the radio waveband than when it goes undetected.
This implies a connection between the radio emission mechanism
and the spin-down torque. The interpretation presented by the authors is
that the radio emission mechanism is only active when sufficient plasma is
present in the magnetosphere, and that this plasma exerts a torque on the
pulsar, spinning it down faster than in the absence of plasma. 
If this is indeed the case, then all observable pulsars should have $n<3$.

\cite{mel97} suggested that the solution to the $n<3$ problem
is related to the angle between the spin and magnetic axes, $\alpha$,
as well as to currents in the magnetosphere. 
Melatos postulates that the magnetosphere can be considered to be split into
two sections, an inner and outer magnetosphere. The division occurs at 
the `vacuum' radius, the location where particles are no longer confined 
to field lines. The inner magnetosphere will then corotate with the neutron
star and can be considered part of the radius of the rotating dipole. However,
since this radius is less than, but comparable in size to, that of the light
cylinder, the dipole can no longer be treated as a point, but has some
finite size. As a result, $2<n<3$ and $n$ approaches 3 as a pulsar ages.
This model is especially attractive
because it provides an explanation for the large scatter in observed values 
of $n$, and provides a prediction for $n$ given measured values of $\nu$, 
$\dot{\nu}$ and $\alpha$. Given the large uncertainties on known values
of $\alpha$, the model roughly agrees with 
measurements of $n$ for PSRs~B1509$-$58,
B0540$-$69 and the Crab pulsar. PSR~J1119$-$6127 does not appear to have a
well determinable $\alpha$. However, following the Melatos model, 
the measured value of $n=2.91\pm0.05$
predicts a range of $10^{\circ} \le \alpha \le 32^{\circ}$ \citep{ck03}. 
Our recent measurement
of $n$ for PSR~J1846$-$0258 allows a prediction of $\alpha = 8.1 -
9.6^{\circ}$ ($95\%$ confidence). At present, there is no reported
radio detection of this source \citep{kmj+96}, however,
were it one day detected, 
radio polarimetric observations could in principle constrain $\alpha$.

\section{Conclusions}
The five very young pulsars with values of $n$ measured via
phase-coherent timing (Table~\ref{tab:index}) show a wide
range of spin properties and behaviors. The glitch behaviour
exhibited by these pulsars is widely varied, ranging from
PSR~B1509$-$58, which has not glitched in 21.3\,yr of continuous
timing observations, to the Crab pulsar, which experiences a glitch
on average, every $\sim2$\,yr. The measured values of $n$ for these
five pulsars span the relatively wide range between $2.140(9)<n<2.91(5)$. 
With the exception of the value of $n=2.91\pm0.05$ for PSR~J1119$-$6127, 
which is nearly compatible with $n=3$, the measured values of $n$ are 
significantly less than 3. The physical cause of the
spin-down of pulsars remains one of the outstanding problems in pulsar
astronomy.

\begin{acknowledgements}
This research made use of data obtained from the High Energy Astrophysics
Science Archive Research Center Online Service, provided by the NASA-Goddard
Space Flight Center. The Molonglo Radio Observatory is operated by the
University of Sydney. The Parkes radio telescope is part of the Australia 
Telescope which is funded by the
Commonwealth of Australia for operation as a National Facility managed
by CSIRO. MAL is an NSERC PSGS-D Fellow. VMK is a Canada Research
Chair. FPG is a NASA Postdoctoral Program (NPP) Fellow. EVG acknowledges
NASA ADP grant ADP04-0000-0069. Funding for this work was provided by
NSERC, FQRNT, CIAR and CFI. Funding has also 
been provided by NASA \rxte\ grants over the course of this study.
\end{acknowledgements}

\bibliographystyle{spmpsci}


\end{document}